\documentclass[10pt,a4paper]{article}
\usepackage{epsfig,amssymb,amsmath}
\newcommand{\be}{\begin{equation}}
\newcommand{\ee}{\end{equation}}
\newcommand{\I}{{\cal I}}
\newcommand{\bpi}{\mbox{\boldmath $\pi$}}
\newcommand{\pauli}{\mbox{\boldmath $\tau$}}
\newcommand{\half}{\frac{1}{2}}
\newcommand{\twelvth}{\frac{1}{12}}
\font\mybb=msbm10 at 11pt
\def\bb#1{\hbox{\mybb#1}}
\def\bR {\bb{R}}

\newcommand{\news}{\setcounter{equation}{0}}
\def\bea{\begin{eqnarray}}
\def\eea{\end{eqnarray}}

\begin{document} 
\title{\vskip -70pt
\begin{flushright}
{\normalsize DAMTP-2011-40} \\
\end{flushright}
\vskip 60pt
{\bf {\LARGE Classical Skyrmions -- Static Solutions and Dynamics}}\\[30pt]}
\author{\bf {\Large N.S. Manton}\footnote{N.S.Manton@damtp.cam.ac.uk} \\[20pt]
Department of Applied Mathematics and Theoretical Physics\\
University of Cambridge\\
Wilberforce Road, Cambridge CB3 0WA, England}

\date{June 2011}
\maketitle
\vskip 60pt
 
\begin{abstract}
Skyrmions with a realistic value of the pion mass parameter are
expected to be quite compact structures, but beyond baryon number $B=8$
only a few examples are known. The largest of these is the cubically
symmetric $B=32$ Skyrmion which is a truncated piece of the Skyrme
crystal. Here it is proposed that many more such Skyrmions 
could be found, without any restriction on the baryon number, as
pieces of the Skyrme crystal. Particular attention is given to the
possibility of reducing $B$ by 1 by chopping a corner off a cubic
crystal chunk.

Nuclei are modelled by Skyrmions with quantised spin and isospin. Here
it is argued that these quantised states can be approximated by
classically spinning Skyrmions. The orientations of the spinning
Skyrmions corresponding to polarised protons, neutrons and deuterons
are identified. Nuclear collisions and the nuclear spin-orbit force
are discussed in terms of classically spinning Skyrmions. Going beyond
the rigid collective motions of Skyrmions, there are spatially
modulated collective motions, which are oscillatory in time. They 
are argued to describe not just vibrational excitations of nuclei, but also
giant resonance states.

A speculative proposal for identifying quarks inside Skyrmions is
briefly discussed. 
\end{abstract}

\newpage

\section{Introduction}\news

In the Skyrme model, nucleons and nuclei are modelled as dynamical
solitons in a nonlinear field theory of pions \cite{Sk,Sk1}. The
conserved topological charge of the solitons is identified with the physically
conserved baryon number, $B$. The most important solutions are the static
solutions of minimal energy for each baryon number, which are known as
Skyrmions \cite{book}. These can be in translational or rotational 
motion, and can rotate in isospace, the internal symmetry space of the 
three pion fields. Quantising these rigid collective motions gives 
states that are identified with nuclei and some of their excited states.  

Skyrmions have a 50-year history \cite{BR}, starting from 
Skyrme's initial construction of the model and its basic solution with 
unit baryon number. In the last ten years, classical Skyrmions with
all baryon numbers up to 22 and a few cases beyond have been constructed 
numerically \cite{BS3c}. An important insight is that the pion mass
term affects the qualitative shape of
Skyrmions for $B=8$ and above \cite{BS10,BS11}. Fewer solutions with massive
pions have been constructed than with massless pions. This is because
there is a very useful technique, the rational map ansatz \cite{HMS},
which is particularly helpful for constructing solutions with massless pions. 

Solutions with a physically realistic pion mass have been found for
baryon numbers including $B=8,12,16$ and $32$. These are clearly made up of 
subunits with baryon number 4, analogous to the structures occuring in
the alpha-particle model
of nuclei \cite{BMS}. The collective rotational degrees of freedom of these
solutions have been quantised, giving good agreement with known states
of the $N=Z$ nuclei $^8{\rm Be}$, $^{12}{\rm C}$ and $^{16}{\rm O}$
\cite{BMSW,Wood}. The 
collective isospin excitations have also been calculated, leading to
isobar states of $^{12}{\rm B}$, $^{12}{\rm Be}$ etc. These match the
experimental states less well but are still promising.

A further well-established solution in the massless pion case is the 
Skyrme crystal \cite{CJJVJ,KSh}. This is a triply-periodic structure with
remarkably low energy per unit baryon number. Its basic cubic
unit cell contains an object called a half-Skyrmion, with half a 
unit of baryon number, that cannot be entirely isolated in space. The crystal 
should persist when the
pion field is massive, but with rather lower symmetry. Skyrmions
with finite but large baryon number should resemble chunks of this
crystal. However, only one Skyrmion, with baryon number 32, has so far
shown clear evidence for this \cite{Ba,BMS}.

The first purpose of this paper is to collect together some ideas and 
qualitative results that may lead to further detailed understanding of Skyrmion
structure beyond $B=8$, especially for massive pions. 
No substantial computations have been carried out. Most of these ideas 
relate to using the truncated Skyrme crystal to construct solutions 
with a wide range of baryon numbers. Partly, they extend some 
observations and results of Silva Lobo and Ward on the layering
structure of the crystal \cite{S-LW}. A basic idea is 
that it is possible to chop off one unit of baryon number from the 
corner of a cubic crystal chunk. We will also argue that some crystal 
chunk solutions arise using the double (or multi-layer) rational 
map ansatz \cite{MP}. Although this multi-layer generalisation is a 
poorer approximation than the original rational map ansatz, it gives 
some qualitative insight. We will claim, as one result, that there is 
a compact, cubically symmetric Skyrmion of baryon number 14, which may
be useful for modelling $^{14}{\rm C}$. 

The second purpose of this paper is to consider the classical
approximation to quantised states of Skyrmions, where the Skyrmion
dynamics is treated classically. This approach should be valid, at least 
for certain nuclear properties. In particular, we will identify the
proton and neutron as classically spinning $B=1$ Skyrmions, with
different internal orientations. It has proved almost impossible so 
far to consider nuclear reactions using quantised Skyrmions, but 
the classical interactions of spinning Skyrmions should be easier to study 
numerically. We discuss some aspects of how such dynamics could be 
interpreted in terms of polarised nuclear scattering. 

With a nucleon approximated by a classical spinning Skyrmion,
we will also be able to discuss the spin-orbit
coupling between such a Skyrmion and a Skyrmion of larger baryon
number. As is well known, the spin-orbit force is vital for
understanding the shell structure of nuclei, but its large strength,
and how exactly it arises, is rather a mystery \cite{HGG}. Investigation of the
spin-orbit force using the Skyrme model has been restricted so far to
the nucleon-nucleon system \cite{RS,ASW}. We will not resolve 
this issue within the Skyrme model but will suggest how it could 
be further investigated.

More generally, we would like to understand some features of Skyrmion
dynamics that go beyond the rigid collective coordinate motion. A
previously considered example of collective behaviour that is not rigid is the
vibrational excitation of a Skyrmion \cite{Walet,BBT,Ba2}. We will 
consider other types of
motion, that when quantised would correspond to a giant resonance of a
nucleus, that is, an excitation where the spatial distribution of 
electric charge oscillates \cite{HGG}.

Our discussion will be in the context of the basic Skyrme model whose
Lagrangian has just three terms, a sigma model term quadratic in 
derivatives, a Skyrme term quartic in derivatives, and a pion
mass term. The model has all kinds of interesting variants. One can
add a sextic term, or terms describing heavier mesons than the
pions \cite{MZ}, and there are novel variants derived from holographic
QCD where the field theory description is initially in four space
dimensions, and reduces to a highly constrained field theory with an
infinite tower of mesons in three dimensions \cite{SS,Sut}. Since 
our discussion will be largely qualitative, many aspects should be relevant 
in all these variant models.  

\section{The Skyrme Model and Skyrmions}\news

The Skyrme model is a Lorentz invariant, nonlinear field theory
of sigma model type, in which the pion fields $\bpi = (\pi_1, \pi_2, \pi_3)$ 
are combined with a field $\sigma$ into an $SU(2)$-valued scalar field
\be
U(x) = \sigma(x){\bf 1} + i \bpi(x) \cdot \pauli \,,
\ee
where $x = (t,{\bf x}) = (x_0,x_1,x_2,x_3)$ and $\pauli$ are the 
Pauli matrices. $\sigma$ is not an independent field as
$\sigma^2 + \bpi \cdot \bpi = 1$, and the sign of $\sigma$ is
determined by continuity and boundary conditions. The Lagrangian 
is 
\be
L=\int \left\{-\frac{1}{2}\mbox{Tr}(R_\mu R^\mu)+\frac{1}{16}
\mbox{Tr}([R_\mu,R_\nu][R^\mu,R^\nu])
-m^2\mbox{Tr}({\bf 1}-U)\right\} \, d^3x \,,
\label{skylag}
\ee
where $R_{\mu}=(\partial_{\mu} U)U^\dagger$ is a current taking values in 
the Lie algebra of $SU(2)$. For a static field $U({\bf x})$, the energy is
\be
E=\int \left\{-\frac{1}{2}\mbox{Tr}(R_iR_i)-\frac{1}{16}
\mbox{Tr}([R_i,R_j][R_i,R_j])+m^2\mbox{Tr}({\bf 1}-U)\right\} \, d^3x\,.
\label{skyenergy}
\ee
The vacuum is $U = {\bf 1}$, so finite energy fields satisfy the
boundary conditions $\sigma \to 1$, $\bpi \to {\bf 0}$ as $|{\bf x}| \to
\infty$. $E$ is invariant under translations and
rotations in $\bR^3$ and also under $SO(3)$ isospin rotations given 
by
\be
U({\bf x})\mapsto {\cal A}U({\bf x}){\cal A}^\dagger, 
\hskip 1cm {\cal A}\in SU(2)\,,
\label{isospin}
\ee
which rotate the pion fields among themselves in isospace. 

The model has a conserved, integer-valued topological charge
$B$, the baryon number. This is the degree of the
map $U: \bR^3 \to SU(2)$, which is well-defined because
$U \to {\bf 1}$ at spatial infinity. $B$ is the integral of the baryon
density
\be
{\cal B} = -\frac{1}{24\pi^2}\epsilon_{ijk}{\rm Tr}(R_iR_jR_k) \,.
\ee
The energy minimisers for each $B$ are 
called Skyrmions, and their energy $E$ is identified with their rest 
mass. It will be convenient to also refer to local minima 
and saddle points of $E$ with nearby energies as Skyrmions. 

In the absence of the final, pion mass term, there is a chiral symmetry 
$U({\bf x}) \mapsto {\cal A}U({\bf x}){\cal A'}$, 
with ${\cal A}$ and ${\cal A}'$ independent elements of $SU(2)$. This
acts on $(\sigma,\bpi)$ by $SO(4)$ internal rotations. The
chiral symmetry is explicitly broken by the mass term, but even in the
massless model it is spontaneously broken by the vacuum 
boundary conditions. We shall discuss the massless pion version of 
the Skyrme model to a considerable extent, as many Skyrmions and 
the Skyrme crystal are best understood in this limit.   

The expressions (\ref{skylag}) and (\ref{skyenergy}) are in ``Skyrme
units'' and $m$ is a dimensionless pion mass parameter. Traditionally,
the energy and length units of the Skyrme model have been calibrated using 
the proton and delta masses, and $m$ has been given a value of 
approximately $0.5$ \cite{AN}, but recent calibrations using heavier 
nuclei and their rotational excitations suggest a larger length unit,
and in compensation (to keep the physical pion mass fixed) a higher
value of $m$, either $m=1$ \cite{BKS,BS11} or $m=1.125$ \cite{MW}. 

In the discussion below we shall refer to Skyrmions with massless
pions, and to Skyrmions with massive pions. In the latter case we mean
solutions with $m$ in the range 0.5 -- 1.5. The
energy and size of a Skyrmion depend to some extent on the precise value
of $m$ in this range, but this dependence is largely cancelled by the 
energy and length calibration, so we will not specify $m$ more precisely.
The Skyrmions with baryon numbers up to $B=6$ are qualitatively rather
similar for massless and massive pions. The Skyrmions with massless 
pions up to $B=8$ are shown in Fig. 1. Their symmetries are quite surprising. 
Several were first discovered numerically \cite{BTC}, but they can now be
rather well understood using the rational map ansatz for
Skyrmions, reviewed in Section 4. For the energies of the Skyrmions 
see \cite{BS3a,BS3b,BS3c}. These energies satisfy the 
Faddeev--Bogomolny energy bound, $E \ge 12\pi^2|B|$ \cite{Fad}, 
but equality is not attained for any field configuration with non-zero 
$B$ \cite{Ma2}. The bound is approximately attained in the massless 
pion case, notably by the Skyrme crystal, but less so when the pions 
are massive.

\begin{figure}[ht]
\begin{center}
\leavevmode
\vskip -0cm
\epsfxsize=12cm\epsffile{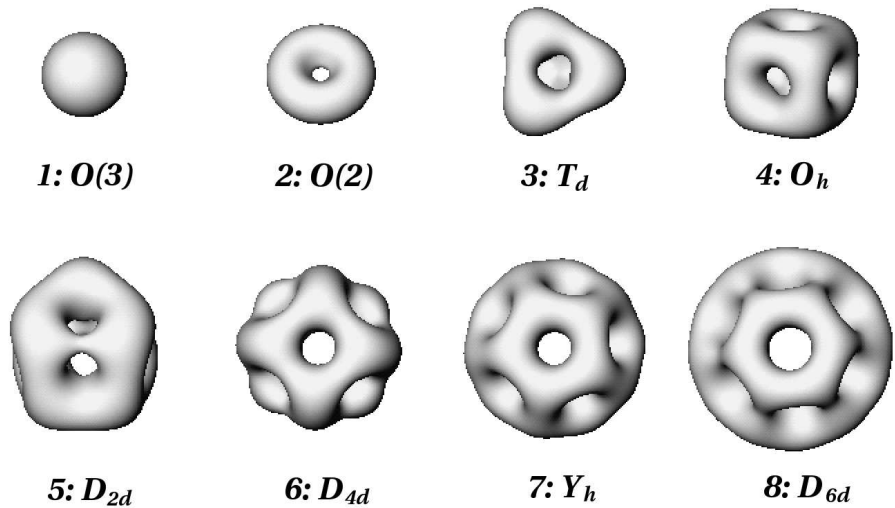}
\caption{Skyrmions for $1\le B\le 8$, with $m=0$. A surface of constant baryon
density is shown, together with the baryon number and symmetry
group. A surface of constant energy density looks similar.}
\label{fig-1-8}
\vskip 0cm
\end{center} 
\end{figure}

The $B=1$ Skyrmion is a spherically symmetric hedgehog, a field 
of the form
\be
U({\bf x}) = \cos f(r){\bf 1} + i\sin f(r) \hat{\bf x} \cdot \pauli
\label{hedgehog}
\ee
where $r = |{\bf x}|$, and $f(0) = \pi$ and $f(\infty) = 0$. Its 
quantised states of spin $\half$ and
isospin $\half$ represent the nucleons \cite{ANW,AN}. 

\begin{figure}[ht]
\begin{center}
\leavevmode
\vskip -0cm
\epsfxsize=6cm\epsffile{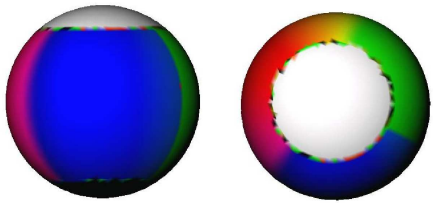}
\caption{$B=1$ Skyrmion (two different views), illustrating the field
  colouring scheme.}
\label{figB=1}
\vskip 0cm
\end{center} 
\end{figure}

The $B=1$ Skyrmion is shown in Fig. 2, using a
colouring scheme based on P.O. Runge's colour sphere. The
colours indicate the value of the normalised pion field
$\hat\bpi$. The white and black points are where $\hat\bpi_3 = \pm1$,
respectively, and $\hat\bpi_1 = \hat\bpi_2 = 0$. The equatorial 
colour circle is where $\hat\bpi_3 =
0$. The primary colours red, blue and green indicate where 
$\hat\bpi_1 + i \hat\bpi_2$ takes the values 
$1, e^{2\pi i/3}, e^{4\pi i/3}$, respectively, and the intermediate points  
$e^{\pi i/3}, -1, e^{5\pi i/3}$ are shown as magenta, cyan and
yellow. The colours are supposed to merge smoothly into each
other as the field $\hat\bpi$ varies. Because of the hedgehog nature
of the $B=1$ Skyrmion, the colouring in Fig. 2 reproduces the colour
sphere itself, but the same colour scheme is used below to illustrate 
a variety of Skyrmions, giving more novel information.

The $B=2$ Skyrmion is a toroidal structure, whose quantisation gives the
deuteron \cite{Kop,BC}, with spin 1 and isospin 0. If this Skyrmion 
remains rigid, its binding 
energy is far greater than that of the deuteron, but it has been shown that
allowing the constituent hedgehog Skyrmions some degree of freedom to
separate gives a spatially larger structure with much lower binding
energy, agreeing much better with the physical deuteron \cite{LMS}. For 
$B=3$ there is tetrahedral symmetry, compatible with the spin states of 
$^{3}{\rm He}$ and $^{3}{\rm H}$ \cite{Ca}. For $B=4$, the Skyrmion has cubic 
symmetry, and its lowest energy quantised state, with spin and isospin
zero, can be identified with the alpha-particle, $^{4}{\rm He}$ \cite{Wal}. 
This Skyrmion is shown in Fig. 3, in a particular orientation in space
and isospace. The cubic symmetry is realised in a
way compatible with the colouring of the colour sphere.

\begin{figure}[ht]
\begin{center}
\leavevmode
\vskip -0cm
\epsfxsize=7cm\epsffile{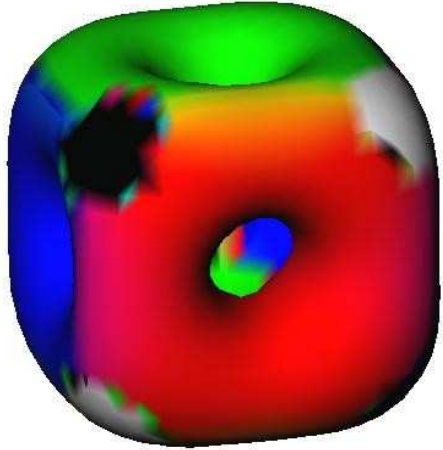}
\caption{$B=4$ Skyrmion.}
\label{figB=4}
\vskip 0cm
\end{center} 
\end{figure}

The $B=4$ solution is particularly
stable. It can be formed by merging four individual hedgehog
Skyrmions on the vertices of a tetrahedron, oriented to
maximally attract. This configuration has tetrahedral symmetry. As the
Skyrmions relax and reach the $B=4$ solution, the symmetry
jumps to cubic, precisely at the moment the energy minimum is attained. 
A consequence is that the $B=4$ Skyrmion can be deformed
equally easily into the initial tetrahedral arrangement of four
hedgehogs, associated with one choice of four alternating vertices of 
a cube, or into a dual tetrahedral arrangement
associated with the other choice. Another consequence is that the 
$B=4$ Skyrmion can be regarded as made up of eight half-Skyrmions on 
the vertices of the cube. These are alternately white
and black, as indicated in the figure. Most of the energy density is 
near these corners, although some is on the twelve edges between them. 
The faces and centre are rather low in energy density.

This merging of $B=1$ Skyrmions, and the resulting half-Skyrmion 
structure, suggests that nuclear forces are almost impossible to 
describe at short distances in terms of the interactions of point 
nucleons of unit baryon number. However, a half-Skyrmion does not 
seem to have anything to do with a quark, which should represent a 
third of a baryon.  

We skip over the $B=5$ solution as the corresponding nuclei 
$^{5}{\rm Li}$ and $^{5}{\rm He}$ are unstable and should, if
anything, be modelled by a dynamical $B=1$ Skyrmion orbiting a $B=4$
core. However, the $B=6$ Skyrmion, shown in Fig. 4, looks sensible. It can be
regarded as a bound structure of a $B=4$ Skyrmion (below) with a $B=2$
Skyrmion (above), both slightly deformed. This matches the
understanding that $^{6}{\rm Li}$ easily dissociates into 
$^{4}{\rm He}$ and a deuteron. It is also consistent with the
observation that in $^{6}{\rm He}$ the two halo neutrons 
orbiting the $^{4}{\rm He}$ core appear to be on the same side \cite{Mul}.

\begin{figure}[ht]
\begin{center}
\leavevmode
\vskip -0cm
\epsfxsize=15cm\epsffile{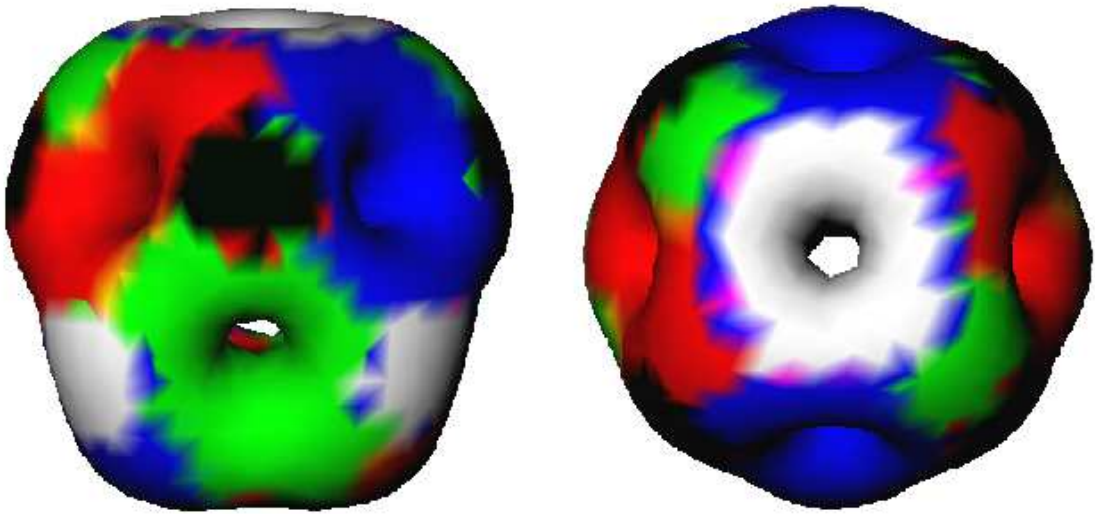}
\caption{$B=6$ Skyrmion (two different views).}
\label{figB=6}
\vskip 0cm
\end{center} 
\end{figure}

The $B=7$ Skyrmion has icosahedral symmetry. This is a
beautiful and surprising result mathematically, and consistent with
the existence of a BPS monopole, a rational map and an $SU(2)$ instanton, all
with topological charge 7 and the same symmetry \cite{book,SiSu}. 
Quantisation of this Skyrmion gives states with isospin $\half$, representing 
$^{7}{\rm Be}$ and $^{7}{\rm Li}$, but the lowest allowed spin is 
$J=\frac{7}{2}$ \cite{Ir,MM}. These nuclei do have states
of quite low energy with spin $\frac{7}{2}$, just 4 MeV above the ground
state for $^{7}{\rm Li}$, but the ground states have spin
$J=\frac{3}{2}$. To accommodate this, one needs a new Skyrmion with a
different symmetry. A plausible candidate has been found, for massive
pions, though its energy value and stability are not confirmed 
\cite{MS}. It is formed by merging
a $B=4$ and a $B=3$ Skyrmion along a common diagonal, preserving $C_3$
symmetry. (A vertex of the $B=4$ cube is inserted into a face of
the $B=3$ tetrahedron.) The resulting $B=7$ Skyrmion has
$D_{3h}$ symmetry, with a $B=1$ Skyrmion located symmetrically between
two $B=3$ clusters. This structure is consistent with the observed 
nature of $^{7}{\rm Li}$ as made of $^{4}{\rm He}$ loosely bound to 
a triton, $^{3}{\rm H}$.

The apparent competition between a more hollow Skyrmion and a more
tightly clustered Skyrmion, at $B=7$, is a precursor to a bifurcation
that more clearly sets in at $B=8$. The Skyrmions with massless pions
remain hollow polyhedral structures up to $B=22$ and
beyond, following the pattern of the $B=8$ solution in Fig. 1. There
is no energetic price to pay for an ever larger interior hole with very small
integrated baryon density, where $U$ is close to $-{\bf 1}$. However 
these hollow Skyrmions fail to match nuclear structure even approximately. For
example, in these Skyrmions, the volumes enclosing significant baryon 
density do not grow linearly with baryon number. For massive pions, 
a region with $U=-{\bf 1}$ has maximal potential energy, so is 
disfavoured by stable
Skyrmions. The Skyrmions instead are more compact structures, and
their volumes grow linearly with baryon number. The known
examples of such Skyrmions are mostly obtained by
binding together several copies of the $B=4$ Skyrmion \cite{BMS},
analogous to the ``molecules'' of alpha-particles that have been used
for decades to model nuclei with $N=Z$ and $B$ a multiple of four. 
The difference from the alpha-particle model
is that the $B=4$ Skyrmions are cubic, rather than pointlike or
spherical, and this affects the geometry of their arrangements 
to some extent. Skyrmions of this type with $B=8, 12, 16$ and $32$ 
have been found. Their shapes are, respectively, linear, an
equilateral triangle, a tetrahedron, and a cube. Variant solutions
with very similar
energies have also been found with different shapes. For example, four $B=4$ 
cubes can be arranged as a flat or bent square, and it is numerically 
very hard to determine which is the true Skyrmion of lowest energy.

The forces that bind $B=1$ Skyrmions together are produced by Yukawa dipoles of
pion fields. This is consistent with the standard one-pion-exchange
force between nucleons. However, in the $B=4$ Skyrmion the dipoles all
cancel out, leaving residual quadrupoles, and the force between $B=4$ cubes is
quadrupole-quadrupole, with an octupole contribution from one of the
pion components. This is very short-range, so the force only operates
when the cubes are close together. The detailed strength of these
quadrupoles is not precisely known. There is a nice agreement in the
alpha-particle model between the binding energy of nuclei up to 
$^{32}{\rm S}$, and the number of alpha-alpha nearest
neighbour bonds \cite{Wef}. The Skyrmion version, based on the
interaction of touching $B=4$ cubes, should be consistent with this.

The force between $B=4$ cubes, all with the same orientation, is strong enough
to create a cubically symmetric arrangement of eight such cubes,
forming a Skyrmion with $B=32$. This is shown in Fig. 5. 
Its structure clearly resembles a 
cubic truncation of the Skyrme crystal, whose crystal symmetry is one of the
cubic space groups. The translational symmetries of the crystal are
approximately realised, locally, in the $B=32$ solution. This Skyrmion
is more symmetric than the arrangements of eight alpha-particles that
are usually considered \cite{Wef}.

\begin{figure}[ht]
\begin{center}
\leavevmode
\vskip -0cm
\epsfxsize=10cm\epsffile{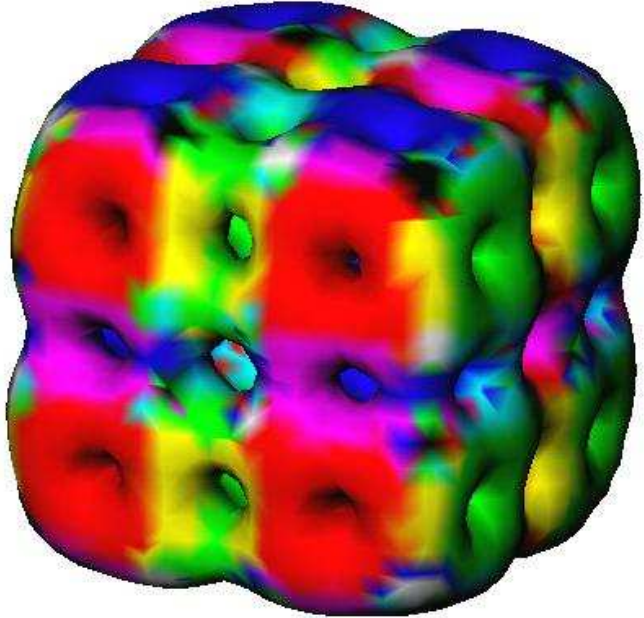}
\caption{$B=32$ Skyrmion.}
\label{figB=32}
\vskip 0cm
\end{center} 
\end{figure}

Below, we will discuss the cubically symmetric $B=32$ Skyrmion in 
more detail, and consider how to subtract or add baryon number to it 
without totally changing its structure. Clearly, by stacking up 27 
$B=4$ cubes, one can obtain a cubically symmetric solution with
$B=108$, and continuing indefinitely, ever larger cubic crystal chunks.
These solutions form a sequence with $8n^3$ half-Skyrmions.
However, in Section 3, we will also argue for the existence of cubically
symmetric chunks of the Skyrme crystal with baryon numbers 1, 14, 63,
etc. These solutions are made from $(2n-1)^3$ half-Skyrmions, 
completed by being surrounded by a pion tail that carries half a unit of baryon
number. Their existence has not been proposed before and we do not know
how stable they are. The $B=1$ example is simply the
usual hedgehog, interpreted as a half-Skyrmion surrounded by a
half-Skyrmion tail. The $B=14$ solution could be important, as the nucleus 
$^{14}{\rm C}$ has recently been recognised as a particularly stable,
essentially spherical nucleus, just as worthy of being interpreted as a
doubly magic nucleus as $^{16}{\rm O}$ \cite{vOer,Bec}. It acts as 
a compact cluster
when interacting with further protons and neutrons to form isotopes
such as $^{18}{\rm O}$. Within the Skyrme model, no phenomenologically
satisfactory solution has been found before with $B=14$. For
massless pions, the hollow polyhedral Skyrmion with $B=14$ is particularly
asymmetric \cite{BS3c}, and for massive pions the only candidate solution is
highly aspherical \cite{BS11}.

\section{Constructing Skyrmions from the Skyrme Crystal}\news

In the massless pion case, the optimal Skyrme crystal structure
was discovered by Castillejo et al. \cite{CJJVJ} and Kugler and 
Shtrikman \cite{KSh}. The crystal has the combined spatial plus internal
symmetries generated by
\bea
(x_1,x_2,x_3)\mapsto(-x_1,x_2,x_3) \,, &\quad&
(\sigma,\pi_1,\pi_2,\pi_3)\mapsto(\sigma,-\pi_1,\pi_2,\pi_3)\,;
\label{cry1}\\
(x_1,x_2,x_3)\mapsto(x_2,x_3,x_1) \,, &\quad&
(\sigma,\pi_1,\pi_2,\pi_3)\mapsto(\sigma,\pi_2,\pi_3,\pi_1)\,;
\label{cry2}\\
(x_1,x_2,x_3)\mapsto(x_1,x_3,-x_2) \,, &\quad&
(\sigma,\pi_1,\pi_2,\pi_3)\mapsto(\sigma,\pi_1,\pi_3,-\pi_2)\,;
\label{cry3}
\eea
and
\be
(x_1,x_2,x_3)\mapsto(x_1+L,x_2,x_3) \,, \quad\quad
 (\sigma,\pi_1,\pi_2,\pi_3)\mapsto(-\sigma,-\pi_1,\pi_2,\pi_3) \,.
\label{cry4}
\ee
Symmetry (\ref{cry1}) is a reflection, (\ref{cry2}) is a $120^\circ$ 
rotation around a diagonal to the three Cartesian axes, 
(\ref{cry3}) is a $90^\circ$ rotation around a Cartesian axis, 
and (\ref{cry4}) is a translation. Note that symmetry (\ref{cry4})
involves a chiral $SO(4)$ rotation, rather than just an $O(3)$
isospace transformation, as in the other symmetries. 

The $\sigma \le 0$ and $\sigma \ge 0$ regions
are perfect cubes of side length $L$, with $\sigma = 0$ on all the
faces. Each cube has similar pion field distributions
and baryon number $\half$. So this is a cubic crystal of half-Skyrmions. 
Numerical calculations, involving either a truncated Fourier series 
expansion of the fields, or a direct construction, estimate 
the energy per baryon to be $E/B=1.036 \times 12\pi^2$ at the optimal
value of $L$. The fields are very well approximated by the 
formulae \cite{CJJVJ}
\bea
\sigma&=& \mp c_1c_2c_3\,, \label{trigcrystal1} \\
\pi_1&=& \pm s_1\sqrt{1-\frac{s_2^2}{2}-\frac{s_3^2}{2}
+\frac{s_2^2s_3^2}{3}}\,,\label{trigcrystal2}
\eea
where $s_i=\sin(\pi x_i/L)$ and $c_i=\cos(\pi x_i/L)$, and $\pi_2$ and
$\pi_3$ are obtained by cyclic permutation of the indices. The sign choice
will be clarified below.

The entire crystal has, of course, infinite baryon number, and is
possibly relevant to neutron stars, but we will not discuss this. 
We are interested instead in how truncated pieces of the
Skyrme crystal -- crystal chunks -- can be used as approximations 
to Skyrmions with the baryon numbers of medium-size and large nuclei. 
Here we know that
the pion mass cannot be neglected, so we really need the Skyrme 
crystal in the case of massive pions. However we shall
assume that the massless pion solution is a good enough approximation,
as it is simpler and better known. The effect of the pion mass term is to
explicitly break chiral symmetry, so the crystal will no longer have
the exact symmetry (\ref{cry4}), but only the two-step symmetry 
\be
(x_1,x_2,x_3)\mapsto(x_1+L,x_2+L,x_3) \,, \quad\quad
(\sigma,\pi_1,\pi_2,\pi_3)\mapsto(\sigma,-\pi_1,-\pi_2,\pi_3) \,.
\ee 
The crystal will also cease to be exactly partitioned into 
half-Skyrmion cubes bounded by surfaces where $\sigma = 0$. However, 
we shall assume that the higher symmetry and structure are
present, even if only approximately.

The simplest type of crystal chunk is one with cubic symmetry. Let us
assume first that the chunk has its centre at the origin. This is
a location in the crystal where the point symmetry group is $O_h$. 
It is also the centre of a half-Skyrmion. The crystal
chunk can be chosen to have its boundaries on the planes 
$x_1 = \pm \frac{2n-1}{2} L$, $x_2 = \pm \frac{2n-1}{2} L$ and 
$x_3 = \pm \frac{2n-1}{2} L$, with $n$ a positive integer. 
Such a chunk has $(2n-1)^3$ half-Skyrmions and total baryon number 
$B=(2n-1)^3/2$. This is a half-integer. The chunk has
$\sigma = 0$ on the entire boundary, and non-zero pion fields taking
all possible values there. This boundary field can be smoothly connected to 
the vacuum at infinity, $\sigma =1$ and $\bpi = {\bf 0}$, by
radial interpolation. This adds a total of half a unit of baryon
number, so these chunks, including their external field, have baryon 
numbers $B=((2n-1)^3 +1)/2 = 4n^3-6n^2+3n$, whose smallest values are 
$1,14,63$ and $172$. These chunks are not yet exact solutions, but we
expect that if they are relaxed they will become Skyrmions retaining a
local crystal structure and at least $O_h$ symmetry. The $B=1$ field 
will relax to the spherically symmetric hedgehog solution, but the $B=14$
solution with $O_h$ symmetry is novel, and no solution with $B=63$ has 
been described before.

The argument that the interpolation adds just $B=\half$ is this. An
individual half-Skyrmion with $U=-{\bf 1}$ at the centre and 
$\sigma =0$ on its boundary is clearly a cubic deformation of 
the central part of a
$B=1$ hedgehog. It can be completed by adding half a unit of baryon
number outside, implying that the contribution of the region that 
extends outwards from each face of the cube is $B=\twelvth$. The
cubic chunk is made up partly of these half-Skyrmions, and partly
of half-Skyrmions with $U={\bf 1}$ at their centres. When the second
type of half-Skyrmion is
completed by outward interpolation from its boundary faces, each face
contributes $B=-\twelvth$, as the completed object has no net baryon
number, the field $\sigma$ being nowhere negative. The same result
comes from noting that the pion fields on the $\sigma =0$ faces have
the opposite orientation from before, which is why the two types of 
half-Skyrmion can meet on a common face. Now a chunk is a
3-dimensional chequer board of these two types of half-Skyrmion, and
should be arranged so that the half-Skyrmions at the chunk corners have
$U=-{\bf 1}$ at their centres, which means choosing the upper signs in
(\ref{trigcrystal1}) and (\ref{trigcrystal2}) if $n$ is odd, and the
lower signs if $n$ is even. Then the number of exposed faces of this 
corner type of half-Skyrmion exceeds the number of the other type by one
on each chunk face, and by six overall, each contributing $B=\twelvth$. 
So the exterior field adds half a unit of baryon number.

Without the sign flips for $n$ even, one would be adding a net $B=-\half$ in
the exterior region. This could create a new type of crystal chunk 
solution, but it is unlikely to be stable, as large regions of
negative baryon density are not energetically favoured in Skyrmions. 

The Skyrmion with $B=14$ which we claim is obtained this way may or may
not be the solution of minimal energy. Previous work found, for
massless pions, a $B=14$ hollow Skyrmion with extremely low symmetry
\cite{BS3c}. A more compact object with high symmetry is likely to be of lower
energy in the massive pion case. We will discuss possible 
constructions of this solution and of the $B=63$ solution in more 
detail in Section 4.

Before doing that we need to consider a cubic crystal chunk with a
different structure \cite{Ba}. Note that in the crystal there are further
locations where the point symmetry group is $O_h$. These are the body
centres of the cubic cells with half-Skyrmions at the vertices. 
Their locations are $(\half L, \half L, \half L)$ and crystal translates
of this. While the point symmetry group is the same, it is realised in
a different way, i.e. the isorotations accompanying rotations are
different. A consequence is that the baryon density is zero at these
locations, whereas it is maximal at the centres of the half-Skyrmions.

To construct a cubic crystal chunk of the new type, with its centre at
the origin, we first translate the crystal by 
$(\half L, \half L, \half L)$. The fields are then well
approximated by the formulae (\ref{trigcrystal1}) and 
(\ref{trigcrystal2}) above, but with 
$s_i, c_i$ replaced by $c_i, -s_i$. However, the resulting field at 
the origin is not $\sigma = \pm 1$, as it needs to be for a cubically 
symmetric crystal chunk to interpolate to the vacuum $\sigma = 1$ at 
infinity. We resolve this by an $SO(4)$ internal rotation. Such
a rotation is not unique, as it can be modified by any $SO(3)$
isorotation. We fix the isospace orientation by requiring 
the field at the half-Skyrmion centres to now be $\pi_3 =\pm 1$, and
requiring that along the $x_3$-axis, $\pi_2$ and $\pi_3$ vanish. 
After this translation and internal rotation, the approximate field
expressions (\ref{trigcrystal1}) and (\ref{trigcrystal2}) become
\bea
{\sigma} &=& \mp \frac{1}{\sqrt{3}}\left(c_1\sqrt{1 -
\frac{c_2^2}{2}-\frac{c_3^2}{2}+\frac{c_2^2c_3^2}{3}}\right. \nonumber\\
&& \left. \quad\quad\quad 
+c_2\sqrt{1-\frac{c_1^2}{2}-\frac{c_3^2}{2}+\frac{c_1^2c_3^2}{3}}
+c_3\sqrt{1-\frac{c_1^2}{2}-\frac{c_2^2}{2}+\frac{c_1^2c_2^2}{3}}\right)
\,,\\
\pi_1&=& \pm \sqrt{\frac{1}{6}}\left(
c_1\sqrt{1-\frac{c_2^2}{2}-\frac{c_3^2}{2}+\frac{c_2^2c_3^2}{3}}
\right. \nonumber\\ 
&& \left. \quad\quad\quad
+c_2\sqrt{1-\frac{c_1^2}{2}-\frac{c_3^2}{2}+\frac{c_1^2c_3^2}{3}}
-2c_3\sqrt{1-\frac{c_1^2}{2}-\frac{c_2^2}{2}+\frac{c_1^2c_2^2}{3}}\right)
\,,\\
\pi_2&=& \pm \frac{1}{\sqrt{2}}  
\left(c_1\sqrt{1-\frac{c_2^2}{2}-\frac{c_3^2}{2}+\frac{c_2^2c_3^2}{3}}
-c_2\sqrt{1-\frac{c_1^2}{2}-\frac{c_3^2}{2}+\frac{c_1^2c_3^2}{3}}\right)
\,,\\
\pi_3&=& \pm s_1s_2s_3\,.
\eea  
The smallest cubic crystal chunk now consists of eight neighbouring 
half-Skyrmion cubes meeting at the origin,
bounded by faces where $\pi_3 =0$. The boundary field can be
interpolated to the vacuum at infinity, and although it is not
obvious, this adds nothing to the baryon number. The general cubic
crystal chunk of this type is bounded by faces at $x_1 = \pm nL$, 
$x_2 = \pm nL$ and $x_3 = \pm nL$. The upper signs in the approximate
formulae for the fields should be chosen for $n$ odd, and the lower 
signs for $n$ even. Such a chunk has $8n^3$ half-Skyrmions and total 
baryon number $B=4n^3$. The notion of half-Skyrmion is a bit different
from before, in terms of the field labelling, but the size, shape,
energy density and baryon density are as before, at least for massless
pions. 

These chunks have baryon numbers $4,32,108,256$ etc., and there is 
more evidence than for the previous type of crystal chunk that they
will relax to Skyrmion solutions. Of these, the 
cubic $B=4$ Skyrmion (Fig. 3), which looks just like a crystal chunk, 
was found independently of the crystal solution. The $B=32$ solution
(Fig. 5) was constructed as a crystal chunk (approximately, and in a different 
orientation) by Baskerville \cite{Ba}, and subsequently by other 
techniques with greater accuracy. 

One feature of the crystal chunks of this second type is that they 
can be regarded as composed of $n^3$ complete $B=4$ cubic Skyrmion subunits, 
all with the same orientation. This should persist for massive pions. 
Fig. 5 illustrates this feature for the case $B=32$, where a $B=4$ subunit is 
visible near each corner of the chunk.\footnote{Remarkably, a variant of this
structure appears, no doubt for a different reason, on a tee-shirt
worn by R.P. Feynman on the cover photo of his book: What Do You Care 
What Other People Think? \cite{Fey}} 
But note that other $B=4$ subunits overlap these, and in 
different internal orientations. For example, there is a subunit
centred at the origin for all $n$. By contrast, the first type of 
crystal chunk has recognisable $B=4$ subunits, but cannot in any 
way be regarded as an aggregate of an integer number of these.

\section{The Rational Map Ansatz}\news

The rational map ansatz is a useful approximate construction of 
Skyrmions, separating the angular from the radial dependence of the 
Skyrme field $U$ \cite{HMS}. The separation is not exact in Skyrmions,
so one should really refer to the rational map approximation rather
than ansatz. 

By stereographic projection, one may define a complex (Riemann sphere) 
coordinate $z = \tan \frac{\theta}{2} \, e^{i\phi}$, where $\theta$ and $\phi$ 
are the usual spherical polar coordinates; alternatively, in terms of
Cartesian coordinates, with $r=|{\bf x}|$, 
\be
z = \frac{x_1 + ix_2}{r + x_3} \,.
\label{zCart}
\ee
The Skyrme field is constructed from a rational function of $z$,
\be
R(z) = \frac{p(z)}{q(z)} \,,
\ee
where $p$ and $q$ are polynomials with no common zero, and from
a radial profile function $f(r)$ satisfying $f(0) = \pi$ and 
$f(\infty) = 0$. By the inverse of (\ref{zCart}), the point $z$
corresponds to the Cartesian unit vector
\be
{\bf n}_z=\frac{1}{1+|z|^2}
(z+\bar z, \, i(\bar z -z), \, 1- |z|^2) \,.
\label{unit1}
\ee
Similarly, an image point $R$ corresponds to the unit vector in isospace
\be
{\bf n}_R = \frac{1}{1 + |R|^2}
(R + \bar{R}, i(\bar{R} - R), 1 - |R|^2) \,.
\ee
The rational map ansatz for the Skyrme field is
\be
U(r,z) = \exp\{if(r){\bf n}_{R(z)}\cdot\pauli\} 
= \cos f(r) {\bf 1} + i\sin f(r){\bf n}_{R(z)}\cdot\pauli \,,
\label{ratansatz}
\ee
generalising the hedgehog formula (\ref{hedgehog}). Note that 
$U=-{\bf 1}$ at the origin. The baryon number $B$
equals the topological degree of the rational map $R:S^2 \to S^2$, 
which is also the algebraic degree of $R$, i.e. the higher of the 
degrees of the polynomials $p$ and $q$. The combination of a
holomorphic map $R$ and a monotonic function $f$ ensures that the
baryon density is everywhere non-negative, which explains why the
rational map ansatz works well.

An $SU(2)$ M\"obius transformation on the domain $S^2$ of the rational map
corresponds to a spatial rotation, whereas an $SU(2)$ M\"obius
transformation on the target $S^2$  corresponds to a rotation of
${\bf n}_R$, and hence to an isorotation of the Skyrme field.
Thus if a rational map $R$ has some symmetry,
then the resulting Skyrme field has that symmetry (i.e.
a spatial rotation can be compensated by an isorotation).

With the rational map ansatz the Skyrme energy 
(\ref{skyenergy}) simplifies to
\be
E=4\pi \int_0^\infty \bigg(
r^2f'^2+2B\sin^2 f(f'^2+1)+\I \, \frac{\sin^4 f}{r^2} + 2m^2r^2(1-\cos f) 
\bigg) \, dr\,,
\label{rmaenergy}
\ee
where $\I$ denotes the angular integral 
\be
\I=\frac{1}{4\pi}\int\bigg(\frac{1+|z|^2}{1+|R|^2}
\bigg\vert\frac{dR}{dz}\bigg\vert\bigg)^4 \frac{2i \, dz d\bar z}
{(1+|z|^2)^2} \,,
\label{i}
\ee
which only depends on the rational map $R(z)$. To minimise the energy (for 
given $B$), it is sufficient to first minimise $\I$ with respect to 
the coefficients occurring in the rational map, and then to solve the
Euler-Lagrange equation for $f(r)$, 
whose coefficients depend on $m,B$ and the minimised $\I$. Optimal 
rational maps have been found for many values
of $B$ \cite{BS3c}, and often have a high degree of symmetry. 
The optimised fields within the rational map ansatz are
good approximations to Skyrmions for massless pions up to $B=22$, 
and used as starting points for numerical 
relaxation to the true Skyrmion solutions, which resemble hollow
polyhedra. The basic ansatz (\ref{ratansatz}) is also useful for
massive pions, but for a much smaller range of $B$. 

The simplest degree 1 rational map is $R(z)=z$, which
is spherically symmetric. The ansatz (\ref{ratansatz}) then
reduces to the hedgehog field (\ref{hedgehog}). 
For $B=2$, 3, 4, 7 the symmetry groups of the numerically computed 
Skyrmions are $O(2),\,T_d,\,O_h,\,Y_h$, respectively. In each 
of these cases there is a rational map with this symmetry, unique up to
rotations and isorotations, namely
\be
R(z)=z^2 \,,\,
R(z)=\frac{z^3-\sqrt{3}iz}{\sqrt{3}iz^2-1}\,,\,
R(z)=\frac{z^4+2\sqrt{3}iz^2+1}{z^4-2\sqrt{3}iz^2+1}\,,\,
R(z)=\frac{z^7-7z^5-7z^2-1}{z^7+7z^5-7z^2+1} \,,
\label{fourmaps}
\ee
and these also minimise $\I$. For $B=5$, 6 and 8, rational maps with
dihedral symmetries are required, and these involve one or two
coefficients that need to be determined numerically. 

The Wronskian of a rational map $R(z) = p(z)/q(z)$ of
degree $B$ is the polynomial
\be
W(z) = p'(z)q(z) - q'(z)p(z)
\ee
of degree $2B-2$. Where $W$ is zero, the derivative $dR/dz$ is
zero, so the baryon density vanishes and the energy density
is low. This explains why the Skyrmions look like polyhedra with holes
in the directions given by the zeros of $W$, and why there are
$2B-2$ such holes, precisely the structures seen in Fig. 1.

For Skyrmions of higher baryon number, and massive pions, it is
helpful to consider generalisations of the rational map ansatz, which
allow for a multi-layered structure. Of these, the first is 
the double rational map ansatz \cite{MP}. This uses two
rational maps $R^{\rm in}(z)$ and 
$R^{\rm out}(z)$, with a profile function $f(r)$ satisfying  
$f(0) = 2\pi$ and $f(\infty) = 0$ and decreasing
monotonically as $r$ increases, passing through $\pi$ at a radius $r_0$.
The ansatz for the Skyrme field is again (\ref{ratansatz}), with
$R(z) = R^{\rm in}(z)$ for $r \le r_0$, and $R(z) = R^{\rm out}(z)$ for 
$r > r_0$. Notice now that $U = {\bf 1}$ 
both at the origin and at spatial infinity, and $U = -{\bf 1}$ on the
entire sphere $r=r_0$. The total baryon number is the sum of the degrees
of the maps $R^{\rm in}$ and $R^{\rm out}$. Optimising this ansatz is
not really worthwhile. Instead the ansatz has been used to
construct initial fields with a given, assumed symmetry, and a
two-layer structure in the radial direction. This is then allowed to
relax numerically, to produce a true Skyrmion with $U = -{\bf 1}$ at
isolated points. 

It is simple to generalise the construction to a $k$-layer rational map
ansatz, using $k$ maps and a profile function satisfying 
$f(0) = k\pi$, $f(\infty) = 0$. Then $U=\pm {\bf 1}$ at the centre,
depending on whether $k$ is even or odd. Experience shows that the
degrees of the maps should increase sharply from the inside to the
outside. In all cases,
the resulting baryon density of the Skyrmion has $2D-2$ holes on its
surface, where $D$ is the degree of the outermost map and $2D-2$ the
degree of its Wronskian. 

A few examples show that the resulting Skyrmion can be a cubic crystal chunk.
In particular, the $B=32$ crystal chunk is obtained by relaxing a
double rational map ansatz, with $B=4$ inside and $B=28$ outside, both
rational maps having cubic symmetry \cite{BMS}. The inner and outer maps are
\bea
R^{\rm in}(z) &=& \frac{p_+}{p_-} \,, \label{Rin32} \\
R^{\rm out}(z) &=& \frac{p_+(ap_+^6 + bp_+^3p_-^3 -p_-^6)}
{p_-(p_+^6 - bp_+^3p_-^3 - ap_-^6)} \,, \label{Rout32}
\eea
where $p_+(z)$ and $p_-(z)$ are the numerator and denominator of the 
rational map of degree 4 in (\ref{fourmaps}), and $a=0.33$ and $b=1.64$ are
determined by numerically minimizing ${\cal I}$.
The outer map has a Wronskian
with 54 zeros, corresponding to nine holes on each cubic face. It
would be interesting to construct the $B=108$ chunk with a three-layer 
rational map ansatz. This would have layers with $B=4$, $B=28$ and
$B=76$. Cubically symmetric rational maps of degree 76 have
rather many parameters, and a suitable map has not previously been
determined. They have 150 Wronskian zeros, which is the right
number of holes, corresponding to six cubic faces each with 25
holes. The map should be chosen so that the inner nine of these holes 
line up with the nine in the face below. 

A suitable map $R(z)$ can probably be constructed as follows. Consider the
outer layer of the crystal chunk as made of half-Skyrmions, located as
in the Skyrme crystal. Since $6^3 - 4^3 = 152$, there are 152 of these 
half-Skyrmions, with alternating values $\pm 1$ of $\pi_3$ at their 
centres (coloured white and black). These pion
field values correspond to $R=0$ and $R=\infty$. So the numerator of
the map should vanish at the 76 points $z$ corresponding to the
locations of the half-Skyrmions with $\pi_3 = 1$, and the
denominator should vanish at the analogous points where  
$\pi_3 = -1$. For example, one white half-Skyrmion is centred at 
$(\frac{3}{2}L, -\frac{1}{2}L, \frac{5}{2}L)$, which corresponds, 
via (\ref{zCart}), to $z = \frac{3-i}{\sqrt{35} + 5}$. Constructing the
numerator and denominator from their zeros is messy but routine. The 
overall constant factor in $R(z)$ is fixed by cubic symmetry. A
simple version of this construction, in terms of eight half-Skyrmions
at the vertices $(\pm \frac{1}{2}L, \pm \frac{1}{2}L, \pm \frac{1}{2}L)$
of a cube, correctly gives the map of degree 4 in
(\ref{fourmaps}) and one could check if the construction
reproduces a degree 28 map of the form (\ref{Rout32}) with similar
coefficients $a$ and $b$ as above.

For the other type of crystal chunk, this algorithm for
constructing the maps contributing to a multi-layer rational map
ansatz fails, as it has the wrong realisation of the $O_h$
symmetry. Nevertheless the double rational map ansatz should be useful 
for constructing the $B=14$ crystal chunk (with massive
pions). Suitable rational maps are presumably
\bea
R^{\rm in}(z) &=& z \,, \label{Rin14} \\
R^{\rm out}(z) &=& \frac{z(z^{12} - (7a+26)z^8 + (6a-39)z^4 + a)}
{az^{12} + (6a-39)z^8 - (7a+26)z^4 + 1} \,, \label{Rout14}
\eea
for some real $a \ne 1$. The outer map, of degree 13, has been 
used in the original rational map ansatz to construct an
$O_h$-symmetric $B=13$ 
Skyrmion (with massless pions); this is not the Skyrmion of lowest 
energy, but has an energy about 1\% higher \cite{BS3c}. The relative 
orientations of $R^{\rm in}$ and $R^{\rm out}$ are chosen carefully 
to maintain $O_h$ symmetry. A slightly indirect way to check this 
is to look at the equation $R^{\rm out}(z) = R^{\rm in}(z)$. This reduces to
\be
z(z^4 - 1)(z^8 + 14z^4 + 1) = 0 \,,
\ee
where the left hand side is a product of two Klein polynomials, the
face and vertex polynomials of a cube. The 14 roots (one of which is
at infinity) correspond to the angular directions where baryon density
is concentrated.

(There was a partly successful previous construction of a $B=14$ 
Skyrmion with this structure, using as $R^{\rm in}$ the degree 7 
map with icosahedral symmetry, and as $R^{\rm out}$ the same degree 
7 map rotated by $90^{\circ}$ \cite{MP}. Together these have
tetrahedral symmetry $T_h$, but the field appears to relax to one with
cubic symmetry. Starting with $B=1$ inside and $B=13$ outside
should be better.)

The crystal chunk with $B=63$ is even less understood. It should be
possible to construct it by relaxing a three-layer rational map 
ansatz, the layers having degrees 1, 13 and 49. However, the degree 49
map is unknown. Its Wronskian will have 96 zeros, corresponding to 
six cubic faces each with 16 holes.

It is not clear whether it is easier to construct crystal chunks
by cutting the infinite crystal suitably, or by building up 
layers using the rational map ansatz. The rational map
approach (even if not used to construct solutions) has the useful
feature that it can be used to calculate the Finkelstein--Rubinstein 
constraints on quantum states \cite{FR}. The relevant calculation for $B=32$ 
is done in \cite{Wood}. 

\section{Chopping Corners off Crystal Chunks}\news

For the Skyrmion solutions with massless pions, up to
$B = 22$, there appears to be no systematic relationship 
between the Skyrmions with neighbouring baryon numbers. In this section 
we will argue that for massive pions, where fewer solutions are
currently known,  
the Skyrmions are possibly more systematically related, at least in
the range from $B=24$ to $B=44$. The proposed Skyrmion structures are 
obtained by locally modifying the crystal chunk structure 
of the $B=32$ Skyrmion. The modification is either by
chopping off one unit of baryon number from a corner or attaching two
units of baryon number to a face. We have no numerical evidence that
the structures obtained are the true minimal energy Skyrmions, but they
should be relatively stable, and even if not minima, could correspond to
classical structures underlying some excited nuclear states of low
spin and their further rotational excitations. Nuclei frequently have
rotational bands built on excited states (see e.g. \cite{Boh} and
references therein).

The key question is this. Is there a $B=31$ Skyrmion that looks
like the $B=32$ cubic Skyrmion, but with one unit of baryon number
removed from a corner, and the structure otherwise little changed? The
answer is unknown, but a piece of evidence in favour is the relation
between the familiar cubic $B=4$ Skyrmion and tetrahedral $B=3$
Skyrmion. 

These latter Skyrmions can be oriented so that the fields near one vertex of
each are rather similar, the only difference being that the angles
between the tetrahedral edges are $60^\circ$, whereas between the
cubic edges they are $90^\circ$. In both cases there is a $C_3$
symmetry around an axis through the vertex. Let us arrange that at 
the vertex $\pi_3 = 1$, with the other field components
vanishing. Opposite this vertex and its neighbouring
faces, the $B=4$ Skyrmion is still quite complicated, with three
further faces and one opposite vertex  where the field
value is $\pi_3 = -1$. The $B=3$ Skyrmion just has a hollow face opposite 
the chosen vertex, where the field value at the centre is $\pi_3 = 1$. 
If one circulates once around this face, then the field circulates 
twice around a loop close to the equatorial circle defined by 
$\pi_1^2 + \pi_2^2 = 1$ (the equatorial colour circle). Such a 
double winding is inevitable when a simple zero of the Wronskian is 
enclosed. Similarly, if one slices the $B=4$ cube orthogonally to the 
diagonal connecting the chosen vertex and the opposite vertex, close 
to mid-way along, then the field also winds twice around the colour 
circle. In summary, one can regard the $B=3$ Skyrmion as the $B=4$ 
Skyrmion with a corner sliced off.

It follows that a mechanism to go from $B=32$ to $B=31$ is to remove
one of the $B=4$ corner subunits and to replace it with 
a $B=3$ Skyrmion with the orientations the same and the tetrahedral vertex
pointing inwards. Three outer faces of the $B=32$ cube will disappear, and 
be replaced by one face orthogonal to a diagonal. One can also think of this
mechanism as slicing off a $B=1$ Skyrmion from the corner. The cubic
symmetry is destroyed, but $C_{3v}$ symmetry remains.

An attempt has been made to carry out this construction by numerically
slicing off part of the crystal chunk and interpolating the fields to
infinity.\footnote{I am grateful to Paul Sutcliffe for trying this.} 
However a problem was that in the $B=32$ Skyrmion, the field
value $U=-{\bf 1}$ occurs at eight groups of four almost coincident
points. A slice tends to include all or none of a group,
and the resulting Skyrmion, after relaxation, has baryon
number 28 or 32. Removing just one $U=-{\bf 1}$ point should be 
possible, but it is necessary to distort the field first, all the while
preserving $C_{3v}$ symmetry.

A net loss of two exterior faces when a Skyrmion loses one unit of
baryon number is what is expected from the multi-layer rational map
ansatz, when the degree of the outer map is decreased by 1. 
We can be more explicit, and describe how to modify the 
outer map $R(z)$ near a cubic vertex at $z_0$, so as to reduce
$B$ by 1. Let us suppose that the map is of degree $D$ and oriented 
(in isospace) so that $R(z_0)=0$. Then $R(z)$ has a linear factor $z - z_0$
in the numerator. Let $z_1, z_2, z_3$ be the locations closest to
$z_0$ where the denominator vanishes (related by the $C_3$ 
symmetry about $z_0$). Then the denominator has a cubic factor 
$(z-z_1)(z-z_2)(z-z_3)$. Now take the limit as $z_1, z_2, z_3$
approach $z_0$. One linear factor cancels between numerator and
denominator, leaving $(z - z_0)^2$ in the denominator. The result is a
map $R^-(z)$ of degree $D-1$, taking the value
$R^-(z_0)=\infty$, and because there is a double pole, the
Wronskian vanishes at $z_0$. This transformation has therefore
reduced the baryon number by 1, and replaced a triplet of distinct 
holes in the outer faces of the cube by a single hole at the location
where there was previously a vertex. The change in the value of the
rational map from 0 to $\infty$ corresponds to the change of sign of 
$\pi_3$ noted when going from the $B=4$ to $B=3$ Skyrmion.  

A further truncation may be desirable in larger crystal chunks. Here
one would modify the numerator of the map $R^-$, so that three
of its zeros approach $z_0$. This would cancel the double zero in the
denominator, leaving a single zero in the numerator. The effect
would be to remove two further units of baryon number from the corner
of the crystal chunk, i.e. three in total. To remove an entire $B=4$
corner subunit will probably require modification of two rational map layers.

If we accept that it is possible to remove a $B=1$ Skyrmion from one
corner of the $B=32$ Skyrmion, then it is likely that 
$B=1$ Skyrmions can be removed from all eight corners, and in any order. These
corners are far enough apart that their fields will not strongly 
influence each other. It is actually possible to remove any number,
such that at least $C_{3v}$ symmetry remains. For
example, two can be removed on diagonally opposite vertices, leaving a
$D_{3d}$ symmetry, or three can be removed from a triangle of vertices
orthogonal to a diagonal. Four can be removed from alternating
vertices, preserving tetrahedral symmetry, $T_d$. This gives a sequence
of at least $C_{3v}$-symmetric Skyrmions for all baryon numbers 
from $B=32$ down to
$B=24$, with cubic symmetry at $B=32$ and $B=24$, and tetrahedral 
symmetry at $B=28$. A cubically symmetric $B=24$ structure matches the
alpha-particle model, where six alpha-particles are optimally arranged
at the vertices of a regular octahedron \cite{Wef}. The Skyrmion
structures for $B=28$ and $B=32$ are more symmetric than in the 
alpha-particle model.

Further evidence for this corner cutting mechanism comes from
considering the hollow $B=13$ Skyrmion mentioned in Section 4, with cubic 
symmetry. This is a solution for massless pions, just slightly greater 
in energy than the solution of minimal energy. We know that there is 
a $B=9$ Skyrmion with tetrahedral symmetry, and a $B=5$ Skyrmion with 
cubic symmetry, both with slightly higher than minimal 
energy \cite{HMS,BS3c}. The tetrahedral and octahedral forms of the
$B=9$ and $B=5$ Skyrmion are just as expected from
slicing four and eight corners off the $B=13$ solution. Slicing corners off a 
cubic $B=14$ Skyrmion is less promising. Here the result would always 
have a $B=1$ Skyrmion at the centre, and this is probably
destabilising when $B$ is reduced to 10 or 6.

Adding baryon number to the $B=32$ cubic Skyrmion is also
possible. Attaching a $B=4$ cubic chunk can be done in many ways but 
creates an object that is best thought of as two loosely bound 
clusters. More stable and symmetric objects are obtained by attaching
two units of baryon number. A rather strongly bound object is
probably produced by attaching a $B=2$ torus to the centre of a face
of the $B=32$ Skyrmion (with the torus parallel to the face). The torus
can be oriented in isospace so that the fields match at the central contact
point and $C_{4v}$ symmetry is preserved. In the resulting $B=34$ 
Skyrmion the attached torus may appear as half of a partially 
embedded $B=4$ cube.    

If this construction is possible, and the result stable, then it
should be possible to repeat it up to six times, attaching tori to
each face. The resulting Skyrmions will have even baryon numbers up 
to $B=44$, the most symmetric being the $B=44$ Skyrmion with cubic symmetry. It
should also still be possible to remove single units of baryon number
from the corners, and in this way create quite a variety of Skyrmions
with all baryon numbers in the range $B=24$ to $B=44$. Some of these
will be quite close to spherical, and others less so. Just a few, with
baryon numbers 24, 32, 36 and 44, will have cubic symmetry, and even
fewer, with baryon numbers 28 and 40, tetrahedral
symmetry. These larger symmetries constrain the possible
quantised spins in a rotational band.

One way to think about the remaining crystal structure of these various
Skyrmions is in terms of layers of half-Skyrmions orthogonal to
the Cartesian axes, although this description starts to break down 
when Skyrmions are cut off from the corners of a cubic crystal
chunk. The infinitely extended version of such a layer is a square
lattice of half-Skyrmions. It is also possible to think in terms of 
layers orthogonal to the principal diagonals of the cube. These layers
are hexagonal lattices of half-Skyrmions, and the way that
neighbouring layers are arranged has
recently been clarified by Silva Lobo and Ward \cite{S-LW}. It is known that 
a truncated piece of a single layer cannot be interpreted, even 
approximately, as a Skyrmion, or as being made of an integer number of 
$B=1$ Skyrmions \cite{BS4}. This is because of wrong boundary conditions above 
or below the layer. But a truncated double-layer structure has been 
found by Battye and Sutcliffe to form Skyrmions \cite{BS11}. Among 
these, the solutions with $B=12$ and maybe $B=18$ have triangular symmetry. 

The possible solutions we mentioned above, in the range $B=24$ to 
$B=32$, with cubic corners cut off, can be thought of as thicker
stacks of truncated layers with triangular symmetry, although 
identifying the precise baryon number in each layer is not particularly clear. 
As an example, consider the $B=32$ Skyrmion itself, made of 64 
half-Skyrmions. Sliced orthogonally to a diagonal, there are layers 
with 1, 3, 6, 10, 12, 12, 10, 6, 3, 1 half-Skyrmions. These layers
consist, alternately, of white and black half-Skyrmions. In particular 
the double layer structure in the middle, with 24 half-Skyrmions and 
baryon number 12, has $D_{3d}$ symmetry and a regular hexagonal 
shape (although the two layers are individually non-regular
hexagons). This appears similar, but not identical, to the $B=12$ 
solution in \cite{BS11}. It would be interesting if a $B=22$ Skyrmion
can be constructed from four triangular layers with 10,12,12,10
half-Skyrmions. 

\section{Spinning Skyrmions}\news

In this section we consider the rigid body dynamics of a
Skyrmion, which is used to model some of the spin and isospin states of 
nuclei with a given baryon number. The standard approach is
that of collective coordinate quantisation. Here one considers a Skyrmion 
and all the related static solutions that differ only by a rotation, a
rotation in isospace, or a translation. These all have the same
energy and none is physically preferred. The space of static solutions
is parametrised by collective coordinates (a point in $\bR^3$ for the
centre of mass, and $SO(3)$ Euler angles for the orientations in space
and isospace). These coordinates are then taken to be time-dependent.
Even though the resulting fields are not exact 
solutions, as no account of Lorentz contraction or centrifugal deformation or
radiation associated with the rotations is considered, the approach
is valid for non-relativistic motion, which is adequate for most of 
nuclear physics. The kinetic terms of the Skyrme Lagrangian
(\ref{skylag}), evaluated for these time-dependent fields, 
give a dynamical system that is essentially a rotor (a rigid body) that can
translate, and rotate in space and isospace. The translational
part is controlled by the Skyrmion rest mass, but for the
rotational motion a more substantial calculation is needed, leading to
numerical expressions for the coupled inertia tensors in both space and 
isospace \cite{BC,BMSW}. When the Skyrmion has symmetries, the inertia
tensors simplify. This dynamical system with its purely kinetic Lagrangian 
is then quantised, following the usual rules for a rigid body. Ignoring
the translational motion, which is rather trivial, one finds spin and
isospin states that can be identified with those of nuclei having the
given baryon number. It is a characteristic feature of the Skyrme
model that one treats in a uniform way both spin excitations and
isospin excitations as quantised collective motions. In conventional nuclear
physics, spin is often treated collectively, leading to rotational
bands, but isospin excitations are usually discussed more
algebraically, with individual nucleons converting from proton to
neutron or vice versa. 

From the Skyrme model we obtain good rotational bands for several
nuclei, e.g. for $^{12}{\rm C}$, and at the same time find spin states for
isobars like $^{12}{\rm B}$ and $^{12}{\rm Be}$ \cite{BMSW}. Typically the
isospin moments of inertia are smaller than the spin moments of
inertia so the isospin excitations require more energy than spin
excitations, in agreement with experimental spectra.

The quantisation procedure is supplemented by symmetry
constraints. When a Skyrmion has discrete symmetries,
involving combined rotations and isorotations, then these symmetries
must leave the quantum state unchanged up to a sign. This significantly cuts
the number of allowed states, and means they do not always lie in
naive rotational bands. The signs are not arbitrary, but determined
topologically, as argued by Finkelstein and Rubinstein \cite{FR}.
For example, a $2\pi$ rotation is always a symmetry, but
acts as $+1$ if $B$ is even and $-1$ if $B$ is odd. The
quantisation of several Skyrmions up to $B=32$ has been carried out,
see e.g. \cite{book,Ir,BMSW,Wood}. The calculation of the FR signs can
be tricky, but has been made easier by an algorithm and formula of 
Krusch \cite{Kr2,Kr3}.

Here we will not discuss any further the quantisation of collective 
coordinates, but discuss instead the classical approximation to 
collective coordinate motion. We will identify the classical 
approximation to a spin-polarised proton or neutron as a classically 
spinning hedgehog Skyrmion. The proton and neutron differ simply
in the way the Skyrmion is oriented as it spins. We will identify
these orientations by noting where the wavefunction of the quantised 
Skyrmion is maximal. We will then consider some other spinning
Skyrmions, for example, the spinning toroidal $B=2$ Skyrmion, and 
identify how one should classically model a polarised deuteron.

We start by describing the classical spinning Skyrmions that
correspond most closely to the spin $\half$ proton and neutron. 
According to Adkins, Nappi and Witten \cite{ANW}, the
wavefunctions of proton and neutron in spin up and spin down states
are simple linear functions of the entries of the $SU(2)$ matrix that controls 
the orientation of the Skyrmion, ${\cal A} =
{\cal A}_0 + i{\cal A}_j\pauli_j$ (where ${\cal A}_0^2 + {\cal A}_1^2
+ {\cal A}_2^2 + {\cal A}_3^2 = 1$). These wavefunctions are
\bea
p^{\uparrow} = \frac{1}{\pi}({\cal A}_1 + i{\cal A}_2) \,, 
&\quad&  n^{\uparrow} = \frac{i}{\pi}({\cal A}_0 + i{\cal A}_3)  \,, \\
p^{\downarrow} = -\frac{i}{\pi}({\cal A}_0 - i{\cal A}_3) \,, 
&\quad&  n^{\downarrow} = -\frac{1}{\pi}({\cal A}_1 - i{\cal A}_2) \,.
\eea
They are not highly localised on the $SU(2)$ 3-sphere, but 
we will treat them as if they were. Each wavefunction has maximal 
magnitude on a great circle of $SU(2)$, and we shall
assume that classically, only the orientations on these circles occur.

Recall that in its standard hedgehog orientation, and with our colour 
scheme, the $B=1$ Skyrmion has its white-black axis along the
$x_3$-axis, with white up and black down,
and is coloured in the $(x_1,x_2)$ plane. In the reversed orientation,
obtained by a rotation by $180^\circ$, it is with black up and
white down.

For a neutron in the spin up state $n^{\uparrow}$, the orientations 
that occur are those obtained from the standard orientation by a 
rotation about the $x_3$-axis. The neutron is spinning with its 
white-black axis fixed, with white up, and the phase variation of
the wavefunction implies that it is spinning anticlockwise
about the positive $x_3$-axis, as expected for a state with positive
spin component relative to this axis. When the neutron is in the spin
down state $n^{\downarrow}$, the orientations that occur are all
those in which the Skyrmion has been rotated by $180^\circ$ about an axis 
in the $(x_1,x_2)$ plane. This means the white-black axis has reversed
direction, and white is down. Again the Skyrmion is spinning, but in
the opposite sense relative to the positive $x_3$-axis. However,
relative to the body fixed axis, that is, the (directed) white-black 
axis, the sense of spin is the same as
before. A neutron state is always one where the spin is anticlockwise
relative to the white-black axis. And this is not restricted to spin
up and spin down states. By taking linear combinations one can obtain a spin
$\half$ state with spin up relative to any oriented axis in space. 
In each case, the white-black axis of the Skyrmion will be parallel to
this spatial axis, and the neutron will spin anticlockwise.

The reader will not be surprised that all this repeats for proton
states, except that the spin is always clockwise with respect to 
the white-black axis. For example, in the spin up proton state 
$p^{\uparrow}$, the spin is anticlockwise relative to the positive 
$x_3$-axis, and therefore clockwise relative to the negative $x_3$-axis. The
Skyrmion orientation is such that the directed white-black axis 
coincides with the negative $x_3$-axis.

The above discussion may appear to imply that a neutron has a
chirality, since it is always spinning anticlockwise. However, this is
illusory, because it depends on the white-black colouring scheme, which
can be reversed.

Let us now turn to the deuteron. The deuteron in the Skyrme model 
is a rotational state of the $B=2$ toroidal Skyrmion, with spin $J=1$ and 
isospin 0. Earlier work established that the axis of rotation is not 
the axis of symmetry, but an axis orthogonal to this \cite{BC,For}. 
The state with $J=1$ and $J_3 = 1$ is easiest to consider. Considered 
classically, it is made up of a polarised proton and neutron. To see
this, we again identify the classical
orientations as those for which the quantum probability is maximal, and
these are where the torus has its axis of symmetry lying in the 
$(x_1,x_2)$ plane. The torus is ``standing on its end''. It is also
spinning anticlockwise about the positive $x_3$-axis, so the symmetry
axis rotates in the $(x_1,x_2)$ plane. Since the isospin is zero, all
possible orientations of the pion fields in isospace occur with equal
probability. Let us pick one, where the field $\pi_3$ takes its
value $1$ at points along both the positive and negative $x_3$-axis. The
spinning $B=2$ Skyrmion can now be interpreted as a pair of $B=1$
Skyrmions, separated in the $x_3$-direction, both spinning
anticlockwise about the positive $x_3$-axis. For the upper Skyrmion 
($x_3 > 0$) the white-black axis is up, whereas for the lower Skyrmion it
is down. Therefore the upper Skyrmion is a neutron with spin up and 
the lower one is a proton with spin up. They have partially merged to form the
torus. This merging occurs where the colours (indicating the phase in
the $(\pi_1, \pi_2)$ plane) agree. For two hedgehog Skyrmions oppositely
oriented, as here, this agreement occurs at a pair of points on opposite 
sides of the $x_3$-axis, rotating steadily as the two Skyrmions spin. This 
is consistent with the picture of a rigidly rotating torus, with 
fixed colouring. Other orientations in isospace are equally likely,
and if white and black are exchanged, the proton will be above the 
neutron. The state of the deuteron with $J=1$ and $J_3 = -1$ is very
similar. The torus, and its individual constituents, are spinning the
opposite way. 

The deuteron in its state with $J=1$ and $J_3 = 0$ is described by 
a Skyrmion where the most likely spatial orientation is with the
symmetry axis coinciding with the $x_3$-axis. The torus is ``lying
down''. As there is no spin component along the $x_3$-axis and no isospin, the
most natural classical description is one where the torus is at
rest. The only recollection that the quantum state has spin 1 is the
fact that not all spatial orientations are equally likely. Here,
unfortunately, it is less clear how to identify the proton and neutron
constituents. One might resolve the ambiguity by thinking of this state as
a quantum superposition of states with $J=1$ and $J_1 = \pm 1$.

The ground states of $^3{\rm He}$ and $^3{\rm H}$, with spin $\half$
and isospin $\half$, are formally rather similar to the states of a
proton and neutron. However the collective coordinates in space and
isospace are independent, and we have not been able to identify if
there is a preferred body-fixed axis of rotation of the tetrahedral $B=3$
Skyrmion, analogous to the white-black axis of the $B=1$ hedgehog. It
would be worthwhile to clarify this.

The alpha-particle has spin and isospin zero. The corresponding ground 
state of the $B=4$ Skyrmion has a wavefunction with no dependence on
the collective coordinates. Classically, one would model this nucleus
by a $B=4$ Skyrmion at rest, with all orientations in space and isospace
equally likely.

As a final example, we note that the ground state of $^6{\rm Li}$
has spin 1 and isospin 0. The $B=6$ Skyrmion has $D_{4d}$ symmetry,
and hence a $C_4$ axis of symmetry. In analogy with the deuteron, the
component of spin along this axis has to vanish. Therefore, a polarised 
$^6{\rm Li}$ nucleus is classically modelled by the $B=6$ Skyrmion spinning
around a body-axis orthogonal to the $C_4$ axis of symmetry. 

Some of the excited states of Skyrmions have zero spin but non-zero
isospin. These should also have an approximate classical
description. The Skyrmion will not physically rotate, but its pion
fields will rotate in isospace. Physical nuclei always have a
well-defined third component of isospin, $I_3$, as this is related to
electric charge $Q$ through the formula $Q = \half B + I_3$. Such
nuclei are modelled by Skyrmions where the fields rotate steadily around the
colour circle, with the white and black points fixed. However the
orientation of the colours relative to the Skyrmion spatial structure
needs to be determined, and this can probably be done using the
quantum states where these are known. It is likely that the
white-black axis in isospace needs to coincide with a
principal axis of the isorotational inertia tensor, in order for
classical motion around this axis to be stationary. The lowest
energy isospin excitations will occur when this axis is the one 
for which the moment of inertia is largest. Detailed examples, like the spin
0, isospin 2 ground state of $^8{\rm He}$, which arises as a quantised
collective state of the $B=8$ Skyrmion, should be investigated. 

\section{Skyrmion-Skyrmion Collisions}\news

Using the classical models of nuclei described in Section 6, we can 
qualitatively describe some nuclear reactions involving, especially, 
polarised nuclei with non-zero spins. Classically one can fix the 
impact parameter, either
at zero for a head-on collision, or at a larger value for a peripheral
collision. A classical description has the best chance of being a good 
approximation when the nuclei are large and moving relatively fast, 
so that the gap in impact parameter between states with orbital 
angular momenta that differ by one unit (i.e. by $\hbar$) is small 
compared with the nuclear radius. This is the regime of heavy ion collisions. 
Also spins should not be small, as spin $\half$ 
states are the least reliably classical. Our examples will 
not satisfy these criteria. Nevertheless we hope they are
illuminating. Calculation of quantised Skyrmion scattering, as a model
for nuclear reactions, was considered by Braaten \cite{Bra}, but is 
so hard that no case has been fully worked
out. With a classical description available, numerical Skyrmion
scattering should be possible, and it will be easier to consider collisions
with a polarised beam and target, than the unpolarised case. 
One might obtain insight into some of the surprising results that have 
been experimentally observed, that are hard to understand with 
conventional two- and three-body nuclear forces \cite{Set,Sek}.

The classical scattering of two $B=1$ Skyrmions has been investigated
previously to some extent, both analytically and numerically
\cite{AKSS,Bra}. Also, numerical calculations of the scattering of 
Skyrmions with higher baryon number have been carried out \cite{BS1}, but with
fixed and rather special choices for the initial orientations. To find
an unpolarised cross section one needs to consider all possible relative
orientations, as well as a range of impact parameters, making the
computations prohibitive. However, we have seen that Skyrmions in
polarised spin states have more precisely defined orientations, so the
modelling of polarised nuclear collisions using spinning Skyrmions
should require less computational effort.
 
Let us first consider proton and neutron collisions with each other,
with beams parallel to the $x_3$-axis. A polarised proton or neutron, 
polarised along the beam axis, has its white-black axis aligned with 
this axis, the direction depending on the polarisation state. 
In a proton-proton collision, where the polarisations relative to 
the positive $x_3$-axis are opposite, or in a proton-neutron collision 
where the polarisations are the same, the collision involves Skyrmions 
approaching with the same colour (either white, or black) at the
closest point of collision. This is an attractive arrangement of the 
Skyrmions, so the nuclear part of the force will be attractive. It is
also the arrangement where in a head-on collision the $B=2$ torus will
be reached. After the collision the torus breaks up in a direction
orthogonal to the beam direction, and there is $90^\circ$ scattering
of the nucleons \cite{Ma3,BS1}. With the other polarisation 
combinations, the colours will be opposite and the force repulsive. 
Scattering angles will be larger in peripheral collisions, and in a
head-on collision the scattering angle will be $180^\circ$. 
This difference will have 
a clear effect on the differential cross section. One might even be
able to model polarisation flip due to the torques exerted by the
colliding Skyrmions. 

Other polarised nuclear collisions can also be modelled using 
Skyrmions, for example, collisions of a polarised deuteron with 
another nucleus. If the
deuteron spin is polarised along the beam axis ($J_3=\pm 1$), then 
it is described by a spinning torus, with its symmetry axis orthogonal 
to the beam and rotating. If the state is polarised with zero spin 
component along the beam ($J_3=0$), then the torus is not rotating, 
but has its symmetry axis along the beam. 

It would be particularly interesting to model deuteron-deuteron 
collisions, by numerically colliding two $B=2$ Skyrmions. For a close 
encounter to occur, the fields at closest approach should match, and 
this controls, to some extent, the relative orientations of the incoming
nuclei. Particularly symmetric would be a collision of two 
deuterons moving parallel to the $x_3$-axis, both with $J_3=0$ 
polarisation. Since deuterons have zero isospin, they do not have a
preferred orientation in isospace. However, to attract at close 
distance, both Skyrmions should be
oriented so that, for example, $\pi_3 = \pm 1$ on the $x_3$-axis, 
with the same value, say $\pi_3 = 1$, on the side closer to the collision
point. Such a collision can reach, with little distortion or twisting,
the $B=4$ cubic Skyrmion. One might be able to estimate the
${\rm d} + {\rm d} \to$ $^{4}{\rm He}$ fusion cross section by 
studying these collisions at various impact parameters. The total 
initial energy of the two incoming Skyrmions (rest energy plus kinetic 
energy) is always greater than the rest energy of the $B=4$ Skyrmion, so
fusion requires dispersion of energy. In a peripheral collision, the
$B=4$ Skyrmion is produced in a spinning state (a quantised $J=4$ state is
allowed theoretically) which could then emit a gamma ray (if one
included electromagnetic fields). In a head-on collision, the $B=4$
Skyrmion is produced in an excited vibrational state, with energy
tending to escape transversely to the collision axis. With enough energy
one would expect to produce a space-star, that is, break-up into four 
$B=1$ Skyrmions moving along axes separated by $90^\circ$ in the centre
of mass frame (see Fig. 9.11 in \cite{book}). This could be a common 
outcome, as two deuterons with a modest amount of kinetic energy 
(required to surmount the Coulomb barrier) have enough energy to
create four nucleons. The Skyrme model suggests that space-star 
production is strongly enhanced in this polarisation state, relative
to others. If break-up does not occur, the vibrational energy could 
be dissipated by electromagnetic emissions.

\section{Further Ideas}\news

Using the classical model for a proton or neutron, we can make
an attempt to model the spin-orbit interaction of a proton or neutron with a
compact larger nucleus. The basic physical consequence of the
spin-orbit force is that it is energetically favourable for spin and 
orbital angular momentum to be aligned, and this effect is strongest 
when the nucleon is close to the nuclear surface \cite{HGG}. It means 
that a nucleon (represented by a Skyrmion) should
roll around the other nucleus. It is not difficult to see that such 
a motion is preferred by the Skyrmion field structure and static
forces that act. This reasoning is
quite different from anything in conventional nuclear physics, as our
Skyrmions are accompanied by a rather rigid classical pion field
configuration. The basic process is best visualised by considering the
$B=1$ Skyrmion shown in Fig. 2 (image on left) rolling around a 
larger Skyrmion, and a good example is the $B=6$ Skyrmion shown 
in Fig. 4 (image on left). For the $B=1$ Skyrmion to be attracted to
the $B=6$ Skyrmion, the field colours should match where the Skyrmions
are closest, as this minimises the field gradient energy. Suppose then
that the $B=1$ Skyrmion is in the standard orientation with white up, and
attached to the $B=6$ Skyrmion on its equator, with, say,
the green colours touching. It is not favourable for the $B=1$ Skyrmion
to roll either upwards or downwards, because black and white regions
repel, but very little potential energy is needed for the $B=1$ Skyrmion 
to roll around the equator. The colour distributions make it clear
that the colours continue to match as the rolling proceeds. This is no
accident, and is a consequence of the baryon density being positive
for both objects. (If the baryon numbers were opposite, the colours
could match in all directions, but then annihilation would occur.) 

In this rolling motion, the spin and orbital motions of the $B=1$ 
Skyrmion are related. Most importantly, the spin is in the
same sense as the orbital angular momentum around the larger object. 
In one orbit around the $B=6$ Skyrmion, the touching colours circulate 
around the colour circle twice, and consequently, the $B=1$ Skyrmion 
spins on its axis three times.

By inspecting the colour schemes of the $B=4$ and $B=32$ Skyrmions,
one can see that there are many tracks around these Skyrmions, bounded by
polygons of white and black points, along which a $B=1$ Skyrmion
could roll. So a spin-orbit force could act on several $B=1$ Skyrmions
simultaneously, leading to a preference for states where they have aligned
spin and orbital angular momentum. This is what nuclear shell model
phenomenology requires.

In Section 6 we discussed the rigid collective motions of Skyrmions. 
Closely related are modulated collective motions of Skyrmions. 
The idea here is to consider a collective motion, but regard it 
as slowly varying across a Skyrmion. An example is a translation 
in the $x_1$-direction whose amplitude is $x_1$-dependent. This is 
a stretching of a Skyrmion, and if time-dependent, leads to a 
vibrational motion. Similarly, a twist about the $x_1$-axis with 
$x_1$-dependent amplitude leads to another type of vibration. Oscillatory 
motion is inevitable when the amplitude is small, because these modulated
collective motions raise the potential energy of a Skyrmion. Break-up
can occur when the amplitude is large. For
a small set of Skyrmions, including those with $B=2,3,4$ and $7$, calculations 
of the vibrational frequencies have been carried out \cite{Walet,BBT,Ba2}. 
Unfortunately, the quantised energies of the excited vibrational states are 
found to be so high that the corresponding nuclei would easily
break up. However the calculations assume that the vibrational motion
is harmonic, which is probably not justified for a typical amplitude
that readily occurs in an excited quantum state. The potential energy
of a typical Skyrmion vibrational mode will flatten out as the
Skyrmion splits into subclusters. Indeed numerically, it was 
found necessary to restrict the classical vibrational amplitudes to be 
very small, which suggests that the quantum states are anharmonic. 
Vibrational motion of larger Skyrmions is likely to be closer to 
harmonic. For example, vibrations of the cubic $B=32$ Skyrmion would 
be worthwhile to study, as the lowest excited states of $^{32}{\rm S}$ 
appear to be vibrational \cite{Ing}.

Another type of modulated collective motion of a Skyrmion is a 
spatially modulated isorotation. This appears not to have
been considered previously. For example, one can isorotate a 
Skyrmion (around some axis in isospace) by an angle that varies
linearly across the Skyrmion, vanishing at the centre. This 
isotwist raises the potential energy, so again leads to an oscillatory 
motion. If the isorotation is around the $\pi_3$-axis -- the
white-black axis -- the effect is to create an oscillating
electric dipole, and since this is a collective motion it corresponds
to a giant dipole resonance. The dipole arises because at any instant
half of the Skyrmion has a clockwise isospin while the other half has
an anticlockwise isospin, which generates local electric charge
densities with opposite signs. Half a period later the motion is
reversed, so the dipole sign is reversed too. In this Skyrmion
description, unlike in the usual picture of a giant resonance in nuclei,
protons and neutrons do not physically move. Only the field
oscillates, consistent with an oscillating flow of charged pions 
across the nucleus. It would be interesting to study this collective 
motion quantitatively for a range of Skyrmions. 

A long-standing challenge in the context of the Skyrme model is to 
identify quark structure inside nucleons and nuclei. The usual
argument is that the Skyrme model is a large $N_c$, low-energy limit
of QCD, where baryons are composed of $N_c$ quarks (with $N_c$
odd) \cite{Wit}. So one should not expect to find three quarks per
baryon. However, there does seem to be a triplet structure in the
colour scheme that is useful for describing Skyrmion
solutions. This might complement the white-black labelling of
half-Skyrmions that has been useful in several ways to keep track of
static Skyrmion structure, and of Skyrmion spin polarisations.

Suppose, within any Skyrmion, we identify the points where $\sigma=0$
and $\hat\bpi$ takes its values corresponding to the primary colours
red, blue and green on the colour circle, as quarks. The quark colours 
(in the QCD sense) can be taken to be
the same as these field colours. Then a Skyrmion of baryon number $B$
will have $B$ quarks of each colour if its baryon density is
everywhere non-negative, because of the topological meaning of the
degree. (Antiquarks can occur if the field values above are at
locations of negative baryon density, but the net quark number is
always $3B$.) These quarks are confined, because a third of a Skyrmion
cannot be spatially isolated.\footnote{A similar
notion of quark structure in a two-dimensional Baby Skyrmion model is
discussed in the recent paper \cite{JSS}.} 

In a $B=1$ hedgehog Skyrmion, the three quarks are always arranged 
at the vertices of an equilateral triangle, and the orientation of the
triangle determines the orientation of the Skyrmion. The quarks
therefore carry useful dynamical information. One could model a
spinning Skyrmion in terms of moving quarks subject to some force law.
Quarks of different colours need to be separated, as the field cannot
have two distinct values at one point, but quarks of the same colour
can coalesce into diquarks. This is seen in the $B=4$ Skyrmion, in its
standard orientation shown in Fig. 3. The red field value occurs with
multiplicity two at two face centres, so the four red quarks
appear as two diquarks. Similarly for blue and green. Other isospace
orientations are possible. For example there could be quarks at the
centre of each edge of the cube. 

This quark idea is not fully worked out, and may be quite
wrong. In particular the quark flavours (up and down) have not been 
identified, nor have the quark spins. They might become clearer by
considering the nucleon and delta quantum states of the $B=1$ Skyrmion.

\section{Conclusions}\news

In this paper we have outlined how the range of known solutions of the
Skyrme model could be extended to encompass a wider range of nuclei
and a wider range of physical phenomena. The idea that medium-size and
larger nuclei should be related to truncated pieces of the Skyrme
crystal, with a realistic positive value for the pion mass parameter, 
is not novel. However, until now, very few Skyrmions have been
found with a clear crystal structure. We have suggested that many more
solutions could be constructed by chopping corners off cubic crystal
chunks, thereby removing one or three units of baryon number, or by adding 
two units of baryon number to a face. It is a challenge to study the
geometry and energetics of this in detail. We have shown how the 
multi-layer rational map ansatz can be helpful both as
a starting point for constructing truncated crystal chunks, and for
giving insight into these structures. We have also indicated how some
relevant maps could be calculated. With more detailed information
on the solutions and their energies, one might be able to 
pinpoint the type of shell structure that is preferred in the Skyrme model.

It is a standard criticism of the Skyrme model that crystalline
structures are too rigid for modelling nuclei, as there is strong
evidence that nucleons within nuclei have significant kinetic energy,
and hence uncertain positions. Despite this, there is also evidence 
that nucleons are strongly correlated, for example into subclusters. 
The Skyrmion crystal structures may be compatible with these
apparently contradictory features of nuclei, because 
they are crystals of half-Skyrmions, and the nucleon 
coordinates are rather ill-defined, if at all. The crystal also has
deuteron and alpha-particle substructures, made of four and eight
half-Skyrmions, respectively. These substructures appear in many
different, overlapping ways. This could resolve the traditional problem
of whether deuterons, alpha-particles etc. are pre-formed or not in 
large nuclei. The Skyrme model says both yes and no.

We have also described in more detail than previously
how certain classically spinning Skyrmions can be identified with
polarised nuclei. In particular, we have identified the orientations
of spinning $B=1$ hedgehog Skyrmions that describe the proton and
neutron, and we have discussed how rolling classical Skyrmions could
explain the spin-orbit interaction between nucleons and larger 
nuclei. Collisions of spinning Skyrmions could be especially effective 
for investigating the outcomes of polarised nuclear collisions, and 
for studying phenenomena like deuteron and alpha-particle knock-out.

Finally we have presented a speculative idea on the
identification of quarks within Skyrmions.

\section*{Acknowledgements}
This work was partly carried out during the author's visits to RIKEN,
Wako, Japan and to the Department of Mathematical Sciences, Durham. The author
thanks Koji Hashimoto for hospitality at RIKEN, and Kimiko Sekiguchi
for discussions and a tour of the RIKEN polarised deuteron facility. He
thanks Paul Sutcliffe, Richard Ward and Jorge Silva Lobo for
discussions in Durham and at the GROUP28 Colloquium at Northumbria
University, Newcastle.

\end{document}